\documentclass[draft,tightenlines,nofootinbib,preprint,aps,eqsecnum,amsmath,amssymb]{revtex4}

\newcommand{\beq}{\begin{equation}}
\newcommand{\eeq}{\end{equation}}
\newcommand{\bea}{\begin{eqnarray}}
\newcommand{\eea}{\end{eqnarray}}

\begin{document}

\title{
General covariance and its implications for Einstein's
space-times.}

\medskip

\author{Luca Lusanna}

\affiliation{ Sezione INFN di Firenze\\ Polo Scientifico\\ Via Sansone 1\\
50019 Sesto Fiorentino (FI), Italy\\ Phone: 0039-055-4572334\\
FAX: 0039-055-4572364\\ E-mail: lusanna@fi.infn.it}

\bigskip
\bigskip

\begin{abstract}

This is a review of the chrono-geometrical structure of special
and general relativity with a special emphasis on the role of
non-inertial frames and of the conventions for the synchronization
of distant clocks. ADM canonical metric and tetrad gravity are
analyzed in a class of space-times suitable  to incorporate
particle physics by using Dirac theory of constraints, which
allows to arrive at a separation of the genuine degrees of freedom
of the gravitational field, the Dirac observables describing
generalized tidal effects, from its gauge variables, describing
generalized inertial effects. A background-independent formulation
(the rest-frame instant form of tetrad gravity) emerges, since the
chosen boundary conditions at spatial infinity imply the existence
of an asymptotic flat metric. By switching off the Newton constant
in presence  of matter this description deparametrizes to the
rest-frame instant form for such matter in the framework of
parametrized Minkowski theories. The problem of the objectivity of
the space-time point-events, implied by Einstein's Hole Argument,
is analyzed.

\bigskip

Talk at the Meeting {\it La Relativita' dal 1905 al 2005: passato,
presente e futuro} organized by SIGRAV and SISM, Department of
Mathematics of the Torino University, June 1, 2005; at ERE2005
{\it A century of relativity physics}, XXVIII Spanish Relativity
Meeting, Oviedo, September 6-10, 2005; at QG05 {\it Constrained
dynamics and quantum gravity 05}, Cala Gonone (Sardinia, Italy),
September 12-16, 2005.

\end{abstract}

\maketitle

\vfill\eject

\section{Introduction}

I will illustrate the status of the understanding of the
chrono-geometrical structure of special and general relativity
from the Hamiltonian point of view. Then I will speak about
Einstein's Hole Argument and of our understanding of the
objectivity of space-time point-events. Finally I will present a
biased list of open problems in this area.

\bigskip

Instead of merging in an enumeration of the main hot problems of
contemporary research like black holes and their entropy, cosmic
censorship and singularity theorems, string and M-theory, loop
quantum gravity and quantum geometry, rotating stars in
astrophysics and gravitational collapse, dark energy, dark matter
and the anisotropy of the cosmic background radiation in the
cosmological context, gravitational lensing, gravitational waves
and their detection, tests on gravity theories from solar system
experiments and binary stars, I will review the basic notions of
relativity emphasizing aspects like the lack of a simultaneity
notion and the importance of a formulation able to give a
well-posed Cauchy problem. Due to general covariance Einstein's
equations are a system of ten partial differential equations,
which cannot be put in normal form. Four of them are not
independent from the others due to Bianchi identities. Other four
are only restrictions on initial data. As a consequence only two
of them contain a genuine dynamical information and eight
components of the 4-metric tensor are left undetermined by
Einstein's equations. Therefore the formulation of their Cauchy
problem is extremely complicated [see Rendall (1998) and Friedrich
and Rendall (2000)for a modern assessment]. However this problem
can be attached in a systematic way in the Hamiltonian approach
based on Dirac theory of constraints and the associated canonical
formulation of the second Noether theorem [see Lusanna (1993)]. In
this way it is possible to develop a strategy for the
determination of the genuine degrees of freedom of the
gravitational field (the Dirac observables, denoted as DO in what
follows), describing its tidal effects, and of a set of hyperbolic
Hamilton equations for their evolution after having fixed all the
arbitrary gauge variables of the gravitational field, describing
its inertial effects. Therefore, a preliminary problem is to find
a well-posed description of special relativistic systems in
non-inertial frames starting from the standard one in inertial
frames dictated by the relativity principle.

\bigskip

This presentation is based  on a theoretical physics viewpoint
aiming to unify gravity and particle physics and to understand how
to reconcile general relativity with quantum theory. I hope to be
able to give a feeling of how {\it heuristic} are most of the
results due to the lack of sufficient mathematical rigor of many
of the tools needed to treat these problems and to stimulate
mathematicians to develop new ideas to refine them.

\section{The Chronogeometrical Structure of Special Relativity}

In the {\it  Annus Mirabilis} 1905 Einstein was able to reconcile
the relativity principle with Maxwell electrodynamics
incorporating Lorentz's partial results and eliminating the
concept of aether. See Norton (2005) for a suggestive
reconstruction of Einstein's line of reasoning to achieve this
result. The outcome was the elimination of Newton absolute time
and of the absolute Euclidean 3-space associated to each instant
of time. According to Newton this instantaneous 3-space had to be
interpreted as a container for matter and its existence amounts to
the philosophical {\it substantivalist} position. This viewpoint
was never accepted by Leibniz, whose {\it relationist} position
refuses the notion of an absolute location of bodies: they are
only defined by their mutual relations with other bodies. The
notion of a container of matter was put in crisis by the advent of
Maxwell electrodynamics, in which fields pervaded the whole
universe, while the relationist point of view influenced Einstein
through the ideas of Mach.

\bigskip

The Galilei group connecting the non-relativistic inertial frames
in accord with the non-relativistic relativity principle was
replaced with the Poincare' group (containing Lorentz
transformations as a subgroup) connecting the relativistic ones
inside the absolute Minkowski space-time according to the
relativistic relativity principle. In both cases Cartesian
coordinates were privileged by this principle. Moreover the
two-way (or round-trip) velocity of light (only one clock is
needed in its definition) was assumed to be $c$, namely {\it
constant} and {\it isotropic} (the light postulates), by Einstein.
The resulting time dilatations and length contraction under
Lorentz transformations became kinematical notions, contrary to
Lorentz's viewpoint according to which they were physical
phenomena, while the light postulates were only a kinematical
convention.

\bigskip

Therefore Einstein's revolution led to unify the independent
notions of time and space in the notion of a 4-dimensional
manifold, the Minkowski space-time, with an absolute (namely
non-dynamical) chrono-geometrical structure. The Lorentz signature
of its 4-metric tensor implies that every time-like observer can
identify the light-cone (the conformal structure, i.e. the locus
of the trajectories of light rays) in each point of the
world-line. But there is {\it no notion of an instantaneous
3-space, of a spatial distance and of a one-way velocity of light
between two observers} (the problem of the synchronization of
distant clocks). Since the relativity principle privileges
inertial observers and Cartesian coordinates $x^{\mu} = (x^o = c
t; \vec x)$ with the time axis centered on them (inertial frames),
the $x^o = const.$ hyper-planes of inertial frames are usually
taken as Euclidean instantaneous 3-spaces, on which all the clocks
are synchronized. Indeed they can be selected with Einstein's
convention for the synchronization of distant clocks to the clock
of an inertial observer. This inertial observer $A$ sends a ray of
light at $x^o_i$ to a second accelerated observer B, who reflects
it towards A. The reflected ray is reabsorbed by the inertial
observer at $x^o_f$. The convention states that the clock of B at
the reflection point must be synchronized with the clock of A when
it signs ${1\over 2}\, (x^o_i + x^o_f)$. This convention selects
the $x^o = const.$ hyper-planes of inertial frames as simultaneity
3-spaces and implies that with this synchronization the two-way
and one-way velocities of light coincide and the spatial distance
between two simultaneous point is the (3-geodesic) Euclidean
distance.

\bigskip

However, real observers are never inertial and for them Einstein's
convention for the synchronization of clocks is not able to
identify globally defined simultaneity 3-surfaces, which could
also be used as Cauchy surfaces for Maxwell equations. The 1+3
{\it point of view} tries to solve this problem starting from the
local properties of an accelerated observer, whose world-line is
assumed to be the time axis of some frame. Since only the observer
4-velocity is given, this only allows to identify the tangent
plane of the vectors orthogonal to this 4-velocity in each point
of the world-line. Then, both in special and general relativity,
this tangent plane is identified with an instantaneous 3-space and
3-geodesic Fermi coordinates are defined on it and used to define
a notion of spatial distance. However this construction leads to
coordinate singularities, because the tangent planes in different
points of the world-line will intersect each other at distances
from the world-line of the order of the (linear and rotational)
{\it acceleration radii} of the  observer (see Mashhoon and Muench
(2002) for their definition). Another type of coordinate
singularity arises in all the proposed uniformly rotating
coordinate systems: if $\omega$ is the constant angular velocity,
then at a distance $r$ from the rotation axis such that $\omega\,
r = c$, the ${}^4g_{oo}$ component of the induced 4-metric
vanishes. This is the so-called {\it horizon problem for the
rotating disk}: the time-like 4-velocity of an observer sitting on
a point of the disk becomes light-like in this coordinate system
when $\omega\, r = c$.

\bigskip

While in particle mechanics one can avoid these problem and
formulate a theory of measurement based on the {\it locality
hypothesis} [standard clocks and rods do not feel acceleration and
at each instant the detectors of the instantaneously comoving
inertial observer give the correct data; see Mashhoon (1990,
2003)], this methodology does not work with continuous media (for
instance the constitutive equations of the electromagnetic field
inside them in non-inertial frames are  unknown) and in presence
of electromagnetic fields when their wavelength is comparable with
the acceleration radii of the observer (the observer is not enough
"static" to be able to measure the frequency of such a wave).
\medskip

See Alba and Lusanna (2003) for a review of these topics.

\bigskip

This state of affairs and the  need of predictability (a
well-posed Cauchy problem for field theory) lead to the necessity
of abandoning the 1+3 point of view and to shift to the 3+1 one.
In this point of view, besides the world-line of an arbitrary
time-like observer, it is given a 3+1 splitting of Minkowski
space-time, namely a foliation of it whose leaves are space-like
hyper-surfaces. Each leaf is both a Cauchy surface for the
description of physical systems and an instantaneous (in general
Riemannian) 3-space, namely a notion of simultaneity implied by a
clock synchronization convention different from Einstein's one.
Even if it is unphysical to give initial data on a non-compact
space-like hyper-surface, this is the only way to be able to use
the existence and uniqueness theorem for the solutions of partial
differential equations. In the more realistic mixed problem, in
which we give initial data on the Earth and we add an arbitrary
information on the null  boundary of the future causal domain of
the Earth (that is we prescribe the data arriving from the rest of
the universe, the ones observed by astronomers), the theorem
cannot be shown to hold!

\bigskip

The extra structure of the 3+1 splitting of Minkowski space-time
allows to enlarge its atlas of 4-coordinate systems with the
definition of {\it Lorentz-scalar observer-dependent radar
4-coordinates} $\sigma^A = (\tau ; \sigma^r)$, $A = \tau , r$.
Here $\tau$ is either the proper time of the accelerated observer
or any monotonically increasing function of it, and is used to
label the simultaneity leaves $\Sigma_{\tau}$ of the foliation. On
each leaf $\Sigma_{\tau}$ the point of intersection with the
world-line of the accelerated observer is taken as the origin of
curvilinear 3-coordinates $\sigma^r$, which can be assumed to be
globally defined since each $\Sigma_{\tau}$ is diffeomorphic to
$R^3$. To the coordinate transformation $x^{\mu} \mapsto \sigma^A$
($x^{\mu}$ are the standard Cartesian coordinates) is associated
an inverse transformation $\sigma^A \mapsto x^{\mu} = z^{\mu}(\tau
, \sigma^r)$, where the functions $z^{\mu}(\tau , \sigma^r)$
describe the embedding of the simultaneity surfaces
$\Sigma_{\tau}$ into Minkowski space-time. The 3+1 splitting leads
to the following induced 4-metric (a functional of the embedding):
${}^4g_{AB}(\tau , \sigma^r) = {{\partial z^{\mu}(\sigma )}\over
{\partial \sigma^A}}\, {}^4\eta_{\mu\nu}\, {{\partial
z^{\nu}(\sigma )}\over {\partial \sigma^B}} = {}^4g_{AB}[z(\sigma
)]$, where ${}^4\eta_{\mu\nu} = \epsilon\, (+ - - -)$ with
$\epsilon = \pm 1$ according to particle physics or general
relativity convention respectively. The quantities
$z^{\mu}_A(\sigma ) = {{\partial z^{\mu}(\sigma )}\over {\partial
\sigma^A}}$ are cotetrad fields on Minkowski space-time.

\bigskip

An admissible 3+1 splitting of Minkowski space-time must have the
embeddings $z^{\mu}(\tau ,\sigma^r)$ of the space-like leaves
$\Sigma_{\tau}$ of the associated foliation satisfying the
M$\o$ller conditions on the coordinate transformation [see
M$\o$ller (1957)]

\begin{eqnarray*}
 && \epsilon\, {}^4g_{\tau\tau}(\sigma ) > 0,\nonumber \\
 &&{}\nonumber \\
 && \epsilon\, {}^4g_{rr}(\sigma ) < 0,\qquad \begin{array}{|ll|} {}^4g_{rr}(\sigma )
 & {}^4g_{rs}(\sigma ) \\ {}^4g_{sr}(\sigma ) & {}^4g_{ss}(\sigma ) \end{array}\, > 0, \qquad
 \epsilon\, det\, [{}^4g_{rs}(\sigma )]\, < 0,\nonumber \\
 &&{}\nonumber \\
 &&\Rightarrow det\, [{}^4g_{AB}(\sigma )]\, < 0.
 \nonumber \\
 \end{eqnarray*}

Moreover, the requirement that the foliation be well defined at
spatial infinity may be satisfied by asking that each simultaneity
surface $\Sigma_{\tau}$ tends to a space-like hyper-plane there,
namely we must have $z^{\mu}(\tau ,\sigma^r)\, \rightarrow\,
x^{\mu}(0) + \epsilon^{\mu}_A\, \sigma^A$ for some set of
orthonormal asymptotic tetrads $\epsilon^{\mu}_A$.

\bigskip

As a consequence, any admissible 3+1 splitting leads to the
definition of a {\it non-inertial frame centered on the given
time-like observer} and coordinatized with Lorentz-scalar
observer-dependent radar 4-coordinates. While  inertial frames
centered on inertial observers are connected by the
transformations of the Poincare' group, the non-inertial ones are
connected by passive frame-preserving diffeomorphism: $\tau
\mapsto \tau^{'}(\tau ,\sigma^r)$, $\sigma^r \mapsto \sigma^{{'}\,
r}(\sigma^s)$. It turns out that M$\o$ller conditions forbid
uniformly rotating non-inertial frames: only differentially
rotating ones are allowed (the ones  used by astrophysicists in
the modern description of rotating stars). In Alba and Lusanna
(2005a) there is a detailed discussion of this topic and there is
the simplest example of  3+1 splittings whose leaves are
space-like hyper-planes carrying admissible differentially
rotating 3-coordinates. Moreover, it is shown that to each
admissible 3+1 splitting are associated two congruences of
time-like observers (the natural ones for the given notion of
simultaneity): i) the Eulerian observers, whose unit 4-velocity
field is the field  of unit normals to the simultaneity surfaces
$\Sigma_{\tau}$; ii) the observers whose unit 4-velocity field is
proportional to the evolution vector field of components $\partial
z^{\mu}(\tau ,\sigma^r)/\partial \tau$: in general this congruence
is non-surface forming having a non-vanishing vorticity (like the
congruence associated to a rotating disk).

\bigskip

The next problem is how to describe physical systems in
non-inertial frames and how to connect different conventions for
clock synchronization. The answer is given by {\it parametrized
Minkowski theories} [see Lusanna (1997, 2004)]. Given any isolated
system (particles, strings, fields, fluids) admitting a Lagrangian
description, one makes the coupling of the system to an external
gravitational field and then replaces the 4-metric
${}^4g_{\mu\nu}(x)$ with the induced metric ${}^4g_{AB}[z(\tau
,\sigma^r)]$ associated to an arbitrary admissible 3+1 splitting.
The Lagrangian now depends not only on the matter configurational
variables but also on the embedding variables $z^{\mu}(\tau
,\sigma^r)$ (whose conjugate canonical momenta are denoted
$\rho_{\mu}(\tau ,\sigma^r)$). Since the action principle turns
out to be invariant under frame-preserving diffeomorphisms, at the
Hamiltonian level there are four first-class constraints ${\cal
H}_{\mu}(\tau ,\sigma^r) = \rho_{\mu}(\tau ,\sigma^r) -
l_{\mu}(\tau ,\sigma^r)\, T^{\tau\tau}(\tau ,\sigma^r) -
z^{\mu}_s(\tau ,\sigma^r)\, T^{\tau s}(\tau ,\sigma^r) \approx 0$
in strong involution with respect to Poisson brackets, $\{ {\cal
H}_{\mu}(\tau ,\sigma^r), {\cal H}_{\nu}(\tau ,\sigma_1^r)\} = 0$.
Here $l_{\mu}(\tau ,\sigma^r)$ are the covariant components of the
unit normal to $\Sigma_{\tau}$, while $z^{\mu}_s(\tau ,\sigma^r)$
are the components of three independent vectors tangent to
$\Sigma_{\tau}$. The quantities $T^{\tau\tau}$ and $T^{\tau s}$
are the components of the energy-momentum tensor of the matter
inside $\Sigma_{\tau}$ describing its energy- and momentum-
densities. As a consequence, Dirac's theory of constraints (or its
geometrical version as presymplectic geometry when only
first-class constraints are present) implies that the
configuration variables $z^{\mu}(\tau ,\sigma^r)$ are arbitrary
{\it gauge variables}. Therefore, all the admissible 3+1
splittings, namely all the admissible conventions for clock
synchronization, and all the admissible non-inertial frames
centered  on time-like observers are {\it gauge equivalent}. By
adding four gauge-fixing constraints $\chi^{\mu}(\tau ,\sigma^r) =
z^{\mu}(\tau ,\sigma^r) - z^{\mu}_M(\tau ,\sigma^r) \approx 0$
[$z^{\mu}_M(\tau ,\sigma^r)$ being an admissible embedding],
satisfying the orbit condition $det\, |\{\chi^{\mu}(\tau
,\sigma^r), {\cal H}_{\nu}(\tau ,\sigma_1^r)| \not= 0$ (implying
the selection of only one point in each gauge orbit inside the
constraint sub-manifold), we identify the description of the
system in the associated inertial frame centered on a given
time-like observer. The resulting effective Hamiltonian for the
$\tau$-evolution turns out to contain the potentials of the {\it
relativistic inertial forces} present in the given non-inertial
frame. Since a  non-inertial frame means the use of its radar
coordinates, we see that already in special relativity {\it
non-inertial Hamiltonians are coordinate-dependent quantities}
like the notion of energy density in general relativity.

\bigskip

As a consequence, the gauge variables $z^{\mu}(\tau ,\sigma^r)$
describe the {\it spatio-temporal appearances} of the phenomena in
non-inertial frames, which, in turn, are associated to extended
physical laboratories using a metrology for their measurements
compatible with the notion of simultaneity of the non-inertial
frame (think to the description of the Earth given by GPS).
Therefore, notwithstanding mathematics tends to use only
coordinate-independent notions, physical metrology forces us to
consider intrinsically coordinate-dependent quantities like the
non-inertial Hamiltonians. For instance, the motion of satellites
around the Earth is governed by a set of empirical coordinates
contained in the software of NASA computers: this is a
metrological standard of space-time around the Earth with a poorly
understood connection with the purely theoretical coordinate
systems. In a few years the European Space Agency will start the
project ACES about the synchronization of a high-precision
laser-cooled atomic clock on the space station with similar clocks
on the Earth  surface by means of microwave signals. If the
accuracy of 5 picosec. will be achieved, it will be possible to
make a coordinate-dependent test of effects at the order $1/c^3$,
like the second order Sagnac effect (sensible to Earth
acceleration) and the general relativistic Shapiro time-delay
created by the geoid. The  two-way velocity of light between an
Earth station and the space station and the synchronization of the
respective clocks are two faces of the same problem.

\bigskip

Inertial frames centered on inertial observers are a special case
of gauge fixing in parametrized Minkowski theories. For each
configuration of an isolated system there is an special 3+1
splitting associated to it: the foliation with space-like
hyper-planes orthogonal to the conserved time-like 4-momentum of
the isolated system. This identifies an intrinsic inertial frame,
the {\it rest-frame}, centered on a suitable inertial observer
(the Fokker-Pryce center of inertia of the isolated system) and
allows to define the {\it Wigner-covariant rest-frame instant form
of dynamics} for every isolated system [see Dirac (1949) for the
various forms of dynamics].

\bigskip

This framework made possible to develop a coherent formalism for
all the aspects of relativistic kinematics both for N particle
systems and continuous bodies and fields [see Alba, Lusanna and
Pauri (2002, 2004)]: i) the classification of the intrinsic
notions of collective variables (canonical non-covariant center of
mass; covariant non-canonical Fokker-Pryce center of inertia;
non-covariant non-canonical M$\o$ller center of energy); ii)
canonical bases of center-of-mass and relative variables; iii)
canonical spin bases and dynamical body-frames for the rotational
kinematics of deformable systems; iv) multipolar expansions for
isolated and open systems; v) the relativistic theory of orbits;
vi) the M$\o$ller radius (a classical unit of length identifying
the region of non-covariance of the canonical center of  mass of a
spinning system around the covariant Fokker-Pryce center of
inertia; it is an effect induced by the Lorentz signature of the
4-metric; it could be used as a physical ultraviolet cutoff in
quantization). See Alba, Lusanna and Pauri (2005) for a
comprehensive review.
\bigskip

Let us remark that in parametrized Minkowski theories a
relativistic particle with world-line $x^{\mu}_i(\tau )$ is
described only by the 3-coordinates $\sigma^r = \eta^r_i(\tau )$
defined by $x^{\mu}_i(\tau ) = z^{\mu}(\tau , \eta^r_i(\tau ))$
and by the conjugate canonical momenta $\kappa_{ir}(\tau )$. The
usual 4-momentum $p_{i\mu}(\tau )$ is a derived quantity
satisfying the mass-shell constraint $\epsilon\, p^2_i = m^2_i$.
Therefore, we have a different description for positive- and
negative- energy particles. All the particles on an admissible
surface $\Sigma_{\tau}$ are simultaneous by construction: this
eliminates the problem of relative times, which for a long time
has been an obstruction to the theory of relativistic bound states
and to relativistic statistical mechanics.

\bigskip

Let us also remark that, differently from Fermi coordinates (a
purely theoretical construction), radar 4-coordinates can be
operationally defined. As shown in Alba and Lusanna (2005a), given
four functions satisfying certain restrictions induced by the
M$\o$ller conditions, the on-board computer of a spacecraft may
establish a grid of radar 4-coordinates in its future.

\bigskip

In Alba and Lusanna (2005b) there is the quantization of
relativistic scalar and spinning particles in a class of
non-inertial frames, whose simultaneity surfaces $\Sigma_{\tau}$
are space-like hyper-planes with arbitrary admissible linear
acceleration and carrying arbitrary admissible differentially
rotating 3-coordinates. It is based on a multi-temporal
quantization scheme for systems with first-class constraints, in
which only the particle degrees of freedom $\eta^r_i(\tau )$,
$\kappa_{ir}(\tau )$ are quantized. The gauge variables,
describing the appearances (inertial effects) of the motion in
non-inertial frames, are treated as c-numbers (like the time in
the Schroedinger equation with a time-dependent Hamiltonian) and
the physical scalar product does not depend on them. The
previously quoted relativistic kinematics has made possible to
separate the center of mass and to verify that the spectra of
relativistic bound states in non-inertial frames are only modified
by inertial effects, being obtained from the inertial ones by
means of a time-dependent unitary transformation. The
non-relativistic limit allows to recover the few existing attempts
of quantization in non-inertial frames as particular cases.

\bigskip

The main open problem is the quantization of the scalar
Klein-Gordon field in non-inertial frames, due to the Torre and
Varadarajan (1999) no-go theorem, according to which in general
the evolution from an initial space-like hyper-surface to a final
one is {\it  not unitary} in the Tomonaga-Schwinger formulation of
quantum field theory. From the 3+1 point of view there is
evolution only among the leaves of an admissible foliation and the
possible way out from the theorem lies in the determination of all
the admissible 3+1 splittings of Minkowski space-time satisfying
the following requirements: i) existence of an instantaneous Fock
space on each simultaneity surface $\Sigma_{\tau}$ (i.e. the
$\Sigma_{\tau}$'s must admit a generalized Fourier transform); ii)
unitary equivalence of the Fock spaces on $\Sigma_{\tau_1}$ and
$\Sigma_{\tau_2}$ belonging to the same foliation (the associated
Bogoliubov transformation must be Hilbert-Schmidt), so that the
non-inertial Hamiltonian is a Hermitean operator; iii) unitary
gauge equivalence of the 3+1 splittings with the Hilbert-Schmidt
property. The overcoming of the no-go theorem would help also in
quantum field theory in curved space-times and in condensed matter
(here the non-unitarity implies non-Hermitean Hamiltonians and
negative energies).

\bigskip

As a final comment, let us note that nearly every relevant
physical system is a field theory with gauge symmetries. This
means that i) we have singular Lagrangian densities whose Hessian
matrix has zero eigenvalues; ii) we must use the second Noether
theorem; iii) we must distinguish between gauge theories
(invariance under a local Lie group acting on an inner space) and
theories with spatio-temporal invariances (invariance under a
group of diffeomorphisms acting also on the space-time); iv) the
Hamiltonian description requires Dirac's theory of constraints and
the physical degrees of freedom are the gauge invariant DO; v) in
gauge theories the gauge variables are redundant variables present
to enforce some kind of manifest covariance, while in theories
with invariances under diffeomorphisms the gauge variables
describe the appearances of phenomena; vi) the only known way to
try to separate DO from gauge variables (namely to separate the
elliptic partial differential equations connected with the
constraints and to arrive to hyperbolic Hamilton equations for the
DO with a well-posed Cauchy problem starting from a set of field
equations restricted by the Noether identities) makes use of
canonical transformations (the Shanmugadhasan ones defined in  the
next Section) in field theory.

Now most of the mathematics used in these steps is not yet
rigorously defined, so that all the results hold only at a
heuristic level.

\section{The Chronogeometrical Structure of General Relativity}

In the years 1913-16 Einstein developed general relativity relying
on the equivalence principle (equality of inertial and
gravitational masses of bodies in free fall). It suggested him the
impossibility to distinguish a constant gravitational field from
the effects of a constant acceleration by  means of local
experiments in sufficiently small regions where the effects of
tidal forces are negligible. This led to the geometrization of the
gravitational interaction and to the replacement of Minkowski
space-time with a pseudo-Riemannian 4-manifold $M^4$ with non
vanishing curvature Riemann tensor. The principle of general
covariance (see Norton (1993) for a review), at the basis of the
tensorial nature of Einstein's equations, has the two following
consequences: i) the invariance of the Hilbert action under {\it
passive} diffeomorphisms (the coordinate transformations in
$M^4$), so that the second Noether theorem implies the existence
of first-class constraints at the Hamiltonian level; ii) the
mapping of solutions of Einstein's equations among themselves
under the action of {\it active} diffeomorphisms of $M^4$ extended
to the tensors over $M^4$ (dynamical symmetries of Einstein's
equations).

\bigskip

The basic field of metric gravity is the 4-metric tensor with
components ${}^4g_{\mu\nu}(x)$ in an arbitrary coordinate system
of $M^4$. The peculiarity of gravity is that the 4-metric field,
differently from the fields of electromagnetic, weak and strong
interactions and from the matter fields, has a {\it double role}:
i) it is the mediator of the gravitational interaction (in analogy
to all the other gauge fields); ii) it determines the
chrono-geometric structure of the space-time $M^4$ in a dynamical
way through the line element $ds^2 = {}^4g_{\mu\nu}(x)\,
dx^{\mu}\, dx^{\nu}$. As a consequence, the gravitational field
{\it teaches relativistic causality} to all the other fields: for
instance it tells to classical rays of light and to quantum
photons and gluons which are the allowed trajectories for massless
particles in each point of $M^4$.

\bigskip

Let us make a comment about the two main existing approaches to
the quantization of gravity.

\medskip

1) Effective quantum field theory and string theory. This approach
contains the standard model of elementary particles and its
extensions. However, since the quantization, namely the definition
of the Fock space, requires a background space-time where it is
possible to define creation and annihilation operators, one must
use the splitting ${}^4g_{\mu\nu} = {}^4\eta^{(B)}_{\mu\nu} +
{}^4h_{\mu\nu}$ and quantize only the perturbation
${}^4h_{\mu\nu}$ of the background 4-metric $\eta^{(B)}_{\mu\nu}$
(usually $B$ is either Minkowski or DeSitter space-time). In this
way property ii) is lost (one uses the fixed non-dynamical
chrono-geometrical structure of the background space-time),
gravity is replaced by a   field of spin two over the background
(and passive diffeomorphisms are replaced by gauge transformations
acting in an inner space) and the only difference among gravitons,
photons and gluons lies in their quantum numbers. The main remnant
of general covariance is the fact that the theory is not
perturbatively renormalizable.

\medskip

2) Loop quantum gravity. This approach never introduces a
background space-time, but being inequivalent to a Fock space, has
problems to incorporate particle physics. It uses a fixed 3+1
splitting of the space-time $M^4$ and it is a quantization of the
associated instantaneous 3-spaces $\Sigma_{\tau}$ (quantum
geometry). However, there is no known way to implement a
consistent unitary evolution (the problem of the super-hamiltonian
constraint) and, since it is usually formulated in spatially
compact space-times without boundary, there is no notion of a
Poincare' group (and therefore no extra dimensions) and a problem
of time (frozen picture without evolution).

\bigskip

For  outside points of view on loop quantum gravity  and string
theory see Nicolai, Peeters and Zamaklar (2005) and Smolin (2003),
respectively.

\bigskip

Let us remark that in all known formulations particle and nuclear
physics are a chapter of the theory of representations of the
Poincare' group in inertial frames in the spatially non-compact
Minkowski space-time. This implies for instance that to speak of
nucleo-synthesis in spatially compact space-times in the
cosmological context is a big extrapolation. \medskip

As a consequence, if one looks at general relativity from the
point of view of particle physics, the main problem to get a
unified theory is how to reconcile the Poincare' group (the
kinematical group of the transformations  connecting inertial
frames) with the diffeomorphism group implying the non-existence
of global inertial frames in general relativity (special
relativity holds only in a small neighborhood of a body in free
fall).

\bigskip

Let us consider the ADM formulation of metric gravity [Arnowitt,
Deser and Misner (1962)] and its extension to tetrad gravity
(needed to describe the coupling of gravity to fermions; it is a
theory of time-like observers endowed with a tetrad field, whose
time-like axis is the unit 4-velocity and whose spatial axes are
associated to a choice of three gyroscopes) obtained by replacing
the ten configurational variables ${}^4g_{\mu\nu}(x)$ with the
sixteen cotetrad fields ${}^4E^{(\alpha )}_{\mu}(x)$ by means of
the decomposition ${}^4g_{\mu\nu}(x) = {}^4E^{(\alpha
)}_{\mu}(x)\, {}^4\eta_{(\alpha )(\beta )}\, {}^4E^{(\beta
)}_{\nu}(x)$ [$(\alpha )$ are flat indices]. Then, after having
restricted the model to globally hyperbolic, topologically
trivial, spatially non-compact space-times (admitting a global
notion  of time), let us introduce a 3+1 splitting of the
space-time $M^4$ and let choose the world-line of a time-like
observer. As in special relativity, let us make a coordinate
transformation to observer-dependent radar 4-coordinates, $x^{\mu}
\mapsto \sigma^A = (\tau ,\sigma^r)$, adapted to the 3+1 splitting
and using the observer world-line as origin of the 3-coordinates.
Again the inverse transformation, $\sigma^A \mapsto x^{\mu} =
z^{\mu}(\tau ,\sigma^r)$, defines the embedding of the leaves
$\Sigma_{\tau}$ into $M^4$. These leaves $\Sigma_{\tau}$ (assumed
to be Riemannian 3-manifolds diffeomorphic to $R^3$, so that they
admit global 3-coordinates $\sigma^r$ and a unique 3-geodesic
joining any pair of points in $\Sigma_{\tau}$) are both Cauchy
surfaces and simultaneity surfaces corresponding to a convention
for clock synchronization. For the induced 4-metric we get

\begin{eqnarray*}
 {}^4g_{AB}(\sigma ) &=& {{\partial z^{\mu}(\sigma )}\over
{\partial \sigma^A}}\, {}^4g_{\mu\nu}(x)\, {{\partial
z^{\nu}(\sigma )}\over
{\partial \sigma^B}} =\nonumber \\
 &=&{}^4E^{(\alpha )}_A\, {}^4\eta_{(\alpha )(\beta )}\,
{}^4E^{(\beta )}_B = \nonumber \\
 &=&\epsilon \left( \begin{array}{cc} (N^2- {}^3g_{rs}\, N^r\, N^s) &
- {}^3g_{su}\, N^u\\ - {}^3g_{ru}\, N^u & -{}^3g _{rs} \end{array}
\right)(\sigma ).\nonumber \\
 \end{eqnarray*}

Here ${}^4E^{(\alpha )}_A(\tau ,\sigma^r)$ are adapted cotetrad
fields, $N(\tau ,\sigma^r)$ and $N^r(\tau ,\sigma^r)$ the lapse
and shift functions and ${}^3g_{rs}(\tau ,\sigma^r)$ the 3-metric
on $\Sigma_{\tau}$ with signature $(+ + +)$. We see that in
general relativity the quantities $z^{\mu}_A = \partial
z^{\mu}/\partial \sigma^A$  are no more cotetrad fields on $M^4$
differently from what happens in special relativity: now they  are
only transition functions between coordinate charts, so that  the
dynamical fields are now the real cotetrad fields ${}^4E^{(\alpha
)}_A(\tau ,\sigma^r)$ and not the embeddings $z^{\mu}(\tau
,\sigma^r)$.

\bigskip

Let us try to identify a class of space-times and an associated
suitable family of admissible 3+1 splittings able to incorporate
particle physics and giving a model for the solar system or our
galaxy (and hopefully allowing an extension to the cosmological
context) with the following further requirements [Lusanna (2001)]:
\medskip

1) $M^4$ must be asymptotically flat at spatial infinity and the
4-metric must  tend asymptotically at spatial infinity to the
Minkowski 4-metric in every coordinate system (this implies that
the 4-diffeomorphisms must tend to the identity at spatial
infinity). Therefore, in these space-times there is an {\it
asymptotic background 4-metric} and this will allow to avoid the
decomposition ${}^4g_{\mu\nu} = {}^4\eta_{\mu\nu} +
{}^4h_{\mu\nu}$ in the bulk.
\medskip

2) The boundary conditions on each leaf $\Sigma_{\tau}$ of the
admissible 3+1 splittings must be such to reduce the Spi group of
asymptotic symmetries [see Wald (1984)] to the ADM Poincare'
group. This means that {\it super-translations}
(direction-dependent quasi Killing vectors, obstruction to the
definition of angular momentum in general relativity) must be
absent, namely that all the fields must tend to their asymptotic
limits in a direction- independent way [see Regge and Teitelboim
(1974) and Beig and O'Murchadha (1987)]. This is possible only if
the admissible 3+1 splittings have all the leaves $\Sigma_{\tau}$
tending to Minkowski space-like hyper-planes orthogonal to the ADM
4-momentum at spatial infinity [Lusanna (2001)]. In turn this
implies that every $\Sigma_{\tau}$ is the rest frame of the
instantaneous 3-universe and that there are asymptotic inertial
observers to be identified with the {\it fixed stars} (in a future
extension to the cosmological context they could be identified
with the privileged observers at rest with respect to the
background cosmic radiation). This requirement implies that the
shift functions vanish at spatial infinity [$N^r(\tau ,\sigma^r)\,
\rightarrow O(1/|\sigma |^\epsilon )$, $\epsilon > 0$, $\sigma^r =
|\sigma |\, {\hat u}^r$], where the lapse function tends to $1$
[$N(\tau ,\sigma^r)\, \rightarrow\, 1 + O(1/|\sigma |^\epsilon )$]
and the 3-metric tends to the Euclidean one [${}^3g_{rs}(\tau
,\sigma^u)\, \rightarrow\, \delta_{rs} + O(1/|\sigma |)$].
\medskip

3) The admissible 3+1 splittings should have the leaves
$\Sigma_{\tau}$ admitting a generalized Fourier transform (namely
they should be Lichnerowicz (1964) 3-manifolds with involution, so
to have the possibility to define instantaneous Fock spaces in a
future attempt of quantization).

\medskip

4) All the fields on $\Sigma_{\tau}$ should belong to suitable
weighted Sobolev spaces, so that $M^4$ has no Killing vectors and
Yang-Mills fields on $\Sigma_{\tau}$ do not present Gribov
ambiguities (due to the presence of gauge symmetries and gauge
copies) [Moncrief (1979), Lusanna (1995), DePietri, Lusanna,
Martucci and Russo (2002)].
\bigskip

In absence of matter the Christodoulou and Klainermann (1993)
space-times are good candidates: they are near Minkowski
space-time in a norm sense, avoid singularity theorems by relaxing
the requirement of conformal completability (so that it is
possible to follow solutions of Einstein's equations on long
times) and admit gravitational radiation at null infinity.

\bigskip

Since the simultaneity leaves $\Sigma_{\tau}$ are the rest frame
of the instantaneous 3-universe,  at the Hamiltonian level it is
possible to define the rest-frame instant form of metric and
tetrad gravity [Lusanna (2001), Lusanna and Russo (2002),
DePietri, Lusanna, Martucci and Russo (2002)]. If matters is
present, the limit of this description for vanishing Newton
constant will produce the rest-frame instant form description of
the same matter in the framework of parametrized Minkowski
theories and the ADM Poincare' generators will tend to the
kinematical Poincare' generators of special relativity. Therefore
we have obtained a model admitting {\it a deparametrization of
general relativity to special relativity}. It is not known whether
the rest-frame condition can be relaxed in general relativity
without having super-translations reappearing, since the answer to
this question is connected with the non-trivial problem of boosts
in general relativity.

\bigskip

Let us now come back to ADM tetrad gravity. The time-like vector
${}^4E^A_{(o)}(\tau ,\sigma^r)$ of the tetrad field
${}^4E^A_{(\alpha )}(\tau ,\sigma^r)$ dual to the cotetrad field
${}^4E^{(\alpha )}_A(\tau ,\sigma^r)$ may be rotated to become the
unit normal to $\Sigma_{\tau}$ in each point by means of a
standard Wigner boost for time-like Poincare' orbits depending on
three parameters $\varphi_{(a)}(\tau ,\sigma^r)$, $a = 1,2,3$:
${}^4E^A_{(o)}(\tau ,\sigma^r) = L^A{}_B(\varphi_{(a)}(\tau
,\sigma^r))\, {}^4{\check E}^B_{(o)}(\tau ,\sigma^r)$. This allows
to define the following cotetrads adapted to the 3+1 splitting
(the so-called {\it Schwinger time gauge}) ${}^4{\check
E}^{(o)}_A(\tau ,\sigma^r) = \Big(N(\tau ,\sigma^r); 0\Big)$,
${}^4{\check E}^{(a)}_A(\tau ,\sigma^r) = \Big(N_{(a)}(\tau
,\sigma^r); {}^3e_{(a)r}(\tau ,\sigma^r)\Big)$, where
${}^3e_{(a)r}(\tau ,\sigma^r)$ are cotriads fields on
$\Sigma_{\tau}$ [tending to $\delta_{(a)r} + O(1/|\sigma |)$ at
spatial infinity] and $N_{(a)} = N^r\, {}^3e_{(a)r}$. As a
consequence, the sixteen cotetrad fields may be replaced by the
fields $\varphi_{(a)}(\tau ,\sigma^r)$, $N(\tau ,\sigma^r)$,
$N_{(a)}(\tau ,\sigma^r)$, ${}^3e_{(a)r}(\tau ,\sigma^r)$, whose
conjugate canonical momenta will be denoted as $\pi_N(\tau
,\sigma^r)$, $\pi_{\vec N\, (a)}(\tau ,\sigma^r)$, $\pi_{\vec
\varphi\, (a)}(\tau ,\sigma^r)$, ${}^3\pi^r_{(a)}(\tau
,\sigma^r)$.

\bigskip

The local invariances of the ADM action imply the existence of 14
first-class constraints (10 primary and 4 secondary): i)
$\pi_N(\tau ,\sigma^r) \approx 0$ implying the secondary
super-hamiltonian constraint ${\cal H}(\tau ,\sigma^r) \approx 0$;
ii) $\pi_{\vec N\, (a)}(\tau ,\sigma^r) \approx 0$ implying the
secondary super-momentum constraints ${\cal H}_{(a)}(\tau
,\sigma^r) \approx 0$; iii) $\pi_{\vec \varphi\, (a)}(\tau
,\sigma^r) \approx 0$; iv) three constraints $M_{(a)}(\tau
,\sigma^r) \approx 0$ generating rotations of the cotriads. As a
consequence there are 14 gauge variables describing the {\it
generalized inertial effects} in the non-inertial frame defined by
the chosen admissible 3+1  splitting of $M^4$ centered on an
arbitrary time-like observer. The remaining independent "two +
two" degrees of freedom are the gauge invariant DO of the
gravitational field describing {\it generalized tidal effects}.
The same degrees  of freedom emerge in ADM metric gravity, where
the configuration variables $N$, $N^r$, ${}^4g_{rs}$ with
conjugate momenta $\pi_N$, $\pi_{\vec N\, r}$, ${}^3\Pi^{rs}$, are
restricted by 8 first-class constraints ($\pi_N(\tau ,\sigma^r)
\approx 0\, \rightarrow {\cal H}(\tau ,\sigma^r) \approx 0$,
$\pi_{\vec N\, r}(\tau ,\sigma^r) \approx 0 \, \rightarrow\, {\cal
H}^r(\tau ,\sigma^r) \approx 0$).
\bigskip

In the canonical approach it is possible to make a  separation of
the gauge variables from the DO by means of a Shanmugadhasan
(1973) canonical transformation [see also Lusanna (1993)]. These
transformations define a canonical basis adapted to the existing
first-class constraints. The constraint presymplectic sub-manifold
defined in phase space by the original first-class constraints is
now defined by the vanishing of an equal number of the new momenta
(Abelianization of the first-class constraints), whose conjugate
configuration variables are the arbitrary gauge variables. The
remaining pairs of the new canonical variables are the DO. While
in finite dimensions the local existence of the Shanmugadhasan
canonical transformations can be demonstrated by using Lie's
theory of function groups and LeviCivita's results about systems
of equations of motion which cannot be  put in normal form, in
field theory the situation is more complicated, because certain
constraints are elliptic partial differential equations. In
function spaces where these equations do not admit zero modes,
these canonical transformations are assumed to exist at least
locally. Dirac (1955) used them to find the DO of the
electromagnetic field: they are the transverse vector potential
${\vec A}_{\perp}$ and the transverse electric field ${\vec
E}_{\perp}$ like in the radiation gauge.

\bigskip

Since no-one knows how to solve the super-hamiltonian constraint
(except that in the post-Newtonian approximation), the best we can
do is to look for a quasi-Shanmugadhasan canonical transformation
adapted to the other 13 first-class constraints (the only
constraints to be Abelianized are $M_{(a)}(\tau ,\sigma^r) \approx
0$ and ${\cal H}_{(a)}(\tau ,\sigma^r) \approx 0$) [DePietri,
Lusanna, Martucci and Russo (2002)]:

\begin{equation}
\begin{minipage}[t]{3cm}
\begin{tabular}{|l|l|l|l|} \hline
$\varphi^{(a)}$ & $N$ & $N_r$ & ${}^3e_{(a)r}$ \\ \hline $\approx
0$ & $\approx 0$ & $  \approx 0 $ & ${}^3{\tilde \pi}^r_{(a)}$
\\ \hline
\end{tabular}
\end{minipage} \hspace{1cm} {\longrightarrow \hspace{.2cm}} \
\begin{minipage}[t]{4 cm}
\begin{tabular}{|lllll|l|l|} \hline
$\varphi^{(a)}$ & $N$ & $N_{(a)}$ & $\alpha_{(a)}$ & $\xi^{r}$ &
$\phi$ & $r_{\bar a}$\\ \hline $\approx0$ &
 $\approx 0$ & $\approx 0$ & $\approx 0$
& $\approx 0$ &
 $\pi_{\phi}$ & $\pi_{\bar a}$ \\ \hline
\end{tabular}
\end{minipage}.
 \nonumber \\
 \end{equation}

Here, $\alpha_{(a)}(\tau ,\sigma^r)$ are three Euler angles and
$\xi^r(\tau ,\sigma^r)$ are three parameters giving a
coordinatization of the action of 3-diffeomorphisms on the
cotriads ${}^3e_{(a)r}(\tau ,\sigma^r)$. The configuration
variable $\phi (\tau ,\sigma^r) = \Big(det\, {}^3g(\tau
,\sigma^r)\Big)^{1/12}$ is the conformal factor of the 3-metric:
it can be shown that it is the unknown in the super-hamiltonian
constraint (also named the Lichnerowicz equation). The gauge
variables are $N$, $N_{(a)}$, $\varphi_{(a)}$, $\alpha_{(a)}$,
$\xi^r$ and $\pi_{\phi}$, while $r_{\bar a}$, $\pi_{\bar a}$,
$\bar a = 1,2$, are the DO of the gravitational field (in general
they are not tensorial quantities).

\bigskip

Even if we do not know the expression of the final variables in
terms of the original ones, we note that this a {\it point}
canonical transformation with known inverse

\beq
 {}^3e_{(a)r}(\tau ,\sigma^u ) =
 {}^3R_{(a)(b)}(\alpha_{(e)}(\tau ,\sigma^u ))\, {{\partial
 \xi^s(\tau ,\sigma^u )}\over {\partial \sigma^r}}\,
\phi^2(\tau , \vec \xi (\tau ,\sigma^u ))\,
 {}^3{\hat e}_{(b)s}( r_{\bar a}
(\tau , \xi^u (\tau ,\sigma^v ))\, ),\nonumber \\
\eeq

\noindent as implied by the study of the gauge transformations
generated by the first-class constraints [${}^3{\hat e}_{(a)r}$
are reduced cotriads, which depend only on the two configurational
DO $r_{\bar a}$].

\medskip

The point nature of the canonical transformation implies that the
old cotriad momenta are linear functionals of the new momenta. The
kernel connecting the old and new momenta satisfy elliptic partial
differential equations implied by i) the canonicity conditions;
ii) the super-momentum constraints ${\cal H}_{(a)}(\tau ,\sigma^r)
\approx 0$; iii) the rotation constraints $M_{(a)}(\tau ,\sigma^r)
\approx 0$.

\bigskip
The first-class constraints are the generators of the Hamiltonian
gauge transformations, under which the ADM action is
quasi-invariant (second Noether theorem):\medskip

i) The gauge transformations generated by the four primary
constraints $\pi_N(\tau ,\sigma^r) \approx 0$, $\pi_{\vec N\,
(a)}(\tau ,\sigma^r) \approx 0$, modify the lapse and shift
functions, namely how densely the simultaneity surfaces are packed
in $M^4$ and which points have the same 3-coordinates on each
$\Sigma_{\tau}$.

ii) Those generated by the three super-momentum constraints ${\cal
H}_{(a)}(\tau ,\sigma^r) \approx 0$ change the 3-coordinates on
$\Sigma_{\tau}$.

iii) Those generated by the super-hamiltonian constraint ${\cal
H}(\tau ,\sigma^r) \approx 0$ transform an admissible 3+1
splitting into another admissible one by realizing a normal
deformation of the simultaneity surfaces $\Sigma_{\tau}$ [see
Teitelboim (1980)]. As a consequence, all the conventions about
clock synchronization are gauge equivalent as in special
relativity.

iv) Those generated by $\pi_{\vec \varphi\, (a)}(\tau ,\sigma^r)
\approx 0$, $M_{(a)}(\tau ,\sigma^r) \approx 0$, change the
cotetrad fields with local Lorentz transformations.

\medskip

In the rest-frame instant form of tetrad gravity there are the
three extra first-class constraints $P^r_{ADM} \approx 0$
(vanishing of the ADM 3-momentum as rest-frame conditions). They
generated gauge transformations which change the time-like
observer whose world-line is used as origin of the 3-coordinates.

\bigskip

Finally let us see which is the Dirac Hamiltonian $H_D$ generating
the $\tau$-evolution in ADM canonical gravity. In spatially
compact space-times without boundary $H_D$ is a linear combination
of the primary constraints (each  one multiplied by an arbitrary
Dirac multiplier, the Hamiltonian version of the undetermined
velocities of the configurational approach whose existence is
implied by the second Noether theorem) plus the secondary
super-hamiltonian and super-momentum constraints multiplied by the
lapse and shift functions respectively (consequence of the
Legendre transform). As a consequence, $H_D \approx 0$ and in the
reduced phase space (quotient of the constraint sub-manifold with
respect to the group of gauge transformations) we get a vanishing
Hamiltonian. This implies the so-called {\it frozen picture} and
the problem of how to reintroduce a temporal evolution. Usually
one considers the normal (time-like) deformation of
$\Sigma_{\tau}$ induced by the super-hamiltonian constraint as an
evolution in a local time variable to be identified (the
multi-fingered time point of view with a local either extrinsic or
intrinsic time): this is the so-called {\it Wheeler-DeWitt
interpretation} (Kuchar (1992,1993) says that the
super-hamiltonian constraint must not be interpreted as a
generator of gauge transformations, but as an effective
Hamiltonian).

\medskip

On the contrary, in spatially non-compact space-times the
definition of functional derivatives and the existence of a
well-posed Hamiltonian action principle (with the possibility of a
good control of the surface terms coming from integration by
parts) require the addition of the {\it DeWitt (1967) surface
term} (living on the surface at spatial infinity) to the
Hamiltonian. It can be shown [Lusanna (2001)] that in the
rest-frame instant form this term, together with a surface term
coming from the Legendre transformation of the ADM action, leads
to the Dirac Hamiltonian

\beq
 H_D = {\check E}_{ADM} + (constraints) =
  E_{ADM} + (constraints) \approx E_{ADM}.\nonumber \\
 \eeq

\noindent Here ${\check E}_{ADM}$ is the {\it strong ADM energy},
a surface term analogous to the one defining the electric charge
as the flux of the electric field through the surface at spatial
infinity in electromagnetism. Since we have ${\check E}_{ADM} =
E_{ADM} + (constraints)$, we see that the non-vanishing part of
the Dirac Hamiltonian is the {\it weak ADM energy} $E_{ADM} = \int
d^3\sigma\, {\cal E}_{ADM}(\tau ,\sigma^r)$, namely the integral
over $\Sigma_{\tau}$ of the ADM energy density (in
electromagnetism this corresponds to the definition of the
electric charge as the volume integral of matter charge density).
Therefore there is no frozen picture but a consistent
$\tau$-evolution.
 \medskip

However, the ADM energy density ${\cal E}_{ADM}(\tau ,\sigma^r)$
is a coordinate-dependent quantity because it depends on the gauge
variables (namely on the inertial effects present in the
non-inertial frame): this is the {\it problem of energy} in
general relativity. Let us remark that in most coordinate systems
${\cal E}_{ADM}(\tau ,\sigma^r)$ does not agree with the
pseudo-energy density defined in terms of the Landau-Lifschiz
pseudo-tensor.

 \bigskip

As a consequence, to get a deterministic evolution for the DO we
must fix the gauge completely, that is we have to add 14
gauge-fixing constraints satisfying an orbit condition (so that
only one point in each gauge orbit inside the constraint
sub-manifold is selected) and to pass to Dirac brackets (the
symplectic structure of the selected copy of the reduced phase
space). The correct way to do it in constraint theory, when there
are secondary constraints, is the following one:\medskip

i) Add a gauge-fixing constraint to the secondary
super-hamiltonian constraint (the choice $\pi_{\phi}(\tau
,\sigma^r) \approx 0$ implies that the DO $r_{\bar a}$, $\pi_{\bar
a}$, remain canonical even if we do not know how to solve this
constraint). This gauge-fixing fixes the form of $\Sigma_{\tau}$,
i.e. the convention for the synchronization of clocks. The
$\tau$-constancy of this gauge-fixing constraint (needed for
consistency) generates a gauge-fixing constraint to the primary
constraint $\pi_N(\tau ,\sigma^r) \approx 0$ for the determination
of the lapse function. The $\tau$-constancy of this new gauge
fixing determines the Dirac multiplier in front of the primary
constraint.

ii) Add three gauge-fixings to the secondary super-momentum
constraints ${\cal H}_{(a)}(\tau ,\sigma^r) \approx 0$. This fixes
the 3-coordinates on each $\Sigma_{\tau}$. The $\tau$-constancy of
these gauge fixings generates the three gauge fixings to the
primary constraints $\pi_{\vec N\, (a)}(\tau ,\sigma^r) \approx 0$
and leads to the determination of the shift functions (i.e. of the
appearances of gravito-magnetism). The $\tau$-constancy of these
new gauge fixings determines the Dirac multipliers in front of the
three primary constraints.

iii) Add six gauge-fixing constraints to the primary constraints
$\pi_{\vec \varphi\, (a)}(\tau ,\sigma^r) \approx 0$,
$M_{(a)}(\tau ,\sigma^r) \approx 0$. This is a fixation of the
cotetrad field which includes a convention on the choice of the
three gyroscopes of every time-like observer of the two
congruences associated to the chosen 3+1 splitting of $M^4$. Their
$\tau$-constancy determines the six Dirac multipliers in front of
these primary constraints.

iv) In the rest-frame instant form we must also add three gauge
fixings to the rest-frame conditions $P^r_{ADM} \approx 0$. The
natural ones are obtained with the requirement that the three ADM
boosts vanish. In this way we select a special time-like observer
as origin of the 3-coordinates (like the Fokker-Pryce center of
inertia in special relativity).

\bigskip

In this way all the gauge variables are fixed to be either
numerical functions or well determined functions of the DO. As a
consequence, in a completely fixed gauge (i.e. in a non-inertial
frame centered on a time-like observer and with its pattern of
inertial forces, corresponding to an extended physical laboratory
with fixed metrological conventions) the ADM energy density ${\cal
E}_{ADM}(\tau ,\sigma^r)$ becomes a well defined function only of
the DO and the Hamilton equations for them with $E_{ADM}$ as
Hamiltonian are a hyperbolic system of partial differential
equations for their determination. For each choice of Cauchy data
for the DO on a $\Sigma_{\tau}$, we obtain a solution of
Einstein's equations in the radar 4-coordinate system associated
to the chosen 3+1 splitting of $M^4$.
\medskip

A universe $M^4$ (a 4-geometry) is the equivalence class of all
the completely fixed gauges with gauge equivalent Cauchy data for
the DO on the associated Cauchy and simultaneity surfaces
$\Sigma_{\tau}$. In each gauge we find the solution for the DO in
that gauge (the tidal effects) and then the explicit form of the
gauge variables (the inertial effects). Moreover, also the
extrinsic curvature of the simultaneity surfaces $\Sigma_{\tau}$
is determined. Since the simultaneity surfaces are asymptotically
flat, it is possible to determine their embeddings $z^{\mu}(\tau
,\sigma^r)$ in $M^4$. As a consequence, differently from special
relativity, the conventions for clock synchronization and the
whole chrono-geometrical structure of $M^4$ (gravito-magnetism,
3-geodesic  spatial distance on $\Sigma_{\tau}$, trajectories of
light rays in each point of $M^4$, one-way velocity of light) are
{\it dynamically determined }.

\bigskip

Let us remark that, if we look at  Minkowski space-time as a
special solution of Einstein's equations with $r_{\bar a}(\tau
,\sigma^r) = \pi_{\bar a}(\tau ,\sigma^r) = 0$ (zero Riemann
tensor, no tidal effects, only inertial effects), we find [Lusanna
(2001)] that the dynamically admissible 3+1 splittings
(non-inertial frames) must have the simultaneity surfaces
$\Sigma_{\tau}$ {\it 3-conformally flat}, because the conditions
$r_{\bar a}(\tau ,\sigma^r) = \pi_{\bar a}(\tau ,\sigma^r) = 0$
imply the vanishing of the Cotton-York tensor of $\Sigma_{\tau}$.
Instead, in special relativity, considered as an autonomous
theory, all the non-inertial frames compatible with the M$\o$ller
conditions are admissible, namely there is much more freedom in
the conventions for clock synchronization.

\bigskip

A first application of this formalism [Agresti, DePietri, Lusanna
and Martucci (2004)]] has been the determination of
post-Minkowskian background-independent gravitational waves in a
completely fixed non-harmonic 3-orthogonal gauge with diagonal
3-metric. It can be shown that the requirements $r_{\bar a}(\tau
,\sigma^r) << 1$, $\pi_{\bar a}(\tau ,\sigma^r) << 1$ lead to a
weak field approximation based on a Hamiltonian linearization
scheme: i) linearize the Lichnerowicz equation, determine the
conformal factor of the 3-metric and then the  lapse and shift
functions; ii) find $E_{ADM}$ in this gauge and disregard all the
terms more than quadratic in the DO; iii) solve the Hamilton
equations for the DO. In this way we get a solution of linearized
Einstein's equations, in which the configurational DO $r_{\bar
a}(\tau ,\sigma^r)$ play the role of the two polarizations of the
gravitational wave and we can evaluate the embedding $z^{\mu}(\tau
,\sigma^r)$ of the simultaneity surfaces of this gauge explicitly.

\section{Einstein's Hole Argument}

In 1914 Einstein (1914), during his researches for developing
general relativity, faced the problem arising from the fact that
the requirement of general covariance would involve a threat to
the physical objectivity of the points of space-time $M^4$, which
in classical field theories are usually assumed to have a well
defined individuality. He formulated the Hole Argument and stated
(our {\it emphasis})

\begin{quotation}
\noindent That this {\it requirement of general covariance}, which
{\it takes away from space and time the last remnant of physical
objectivity}, is a natural one, will be seen from the following
reflexion... (Einstein, 1916, p.117)
\end{quotation}

\bigskip

Assume that $M^4$ contains a {\it hole} ${\cal H}$, that is an
open region where all the non-gravitational fields vanish. It  is
implicitly assumed that the Cauchy surface for Einstein's
equations lies outside ${\cal H}$. Let us consider an active
diffeomorphism $A$ which re-maps the points inside ${\cal H}$, but
is the identity outside ${\cal H}$. For any point $x \in {\cal H}$
we have $x \mapsto D_A\, x \in {\cal H}$. The induced active
diffeomorphism on the 4-metric tensor ${}^4g$, solution of
Einstein's equations, will map it into another solution $D^*_A\,
{}^4g$ ($D^*_A$ is a dynamical symmetry of Einstein's equations)
defined by $D^*_A\, {}^4g(D_A\, x) = {}^4g(x) \not= D^*_A\,
{}^4g(x)$. As  a consequence, we get two solutions of Einstein's
equations with the same Cauchy data outside ${\cal H}$ and it is
not clear how to save the identification of the mathematical
points of $M^4$.

\bigskip

Einstein avoided the problem with the pragmatic {\it
point-coincidence argument}: the only real world-occurrences are
the (coordinate-independent) space-time coincidences (like the
intersection of two world-lines). However, the problem was
reopened by Stachel (1980) and then by Earman and Norton (1987)
and this opened a rich philosophical debate that is still alive
today.

\bigskip

If we insist on the reality of space-time mathematical points
independently from the presence of any physical field (the {\it
substantivalist} point of view in philosophy of science), we are
in trouble with predictability. If we say that ${}^4g$ and
$D^*_A\, {}^4g$ describe the same universe (the so-called {\it
Leibniz equivalence}), we loose any physical objectivity of the
space-time points (the {\it relationist} point of view). Stachel
(1980) suggested that a physical individuation of the point-events
of $M^4$ could be done only by using {\it four individuating
fields depending on the 4-metric on $M^4$}, namely that a tensor
field on $M^4$ is needed to identify the points of $M^4$.
\medskip

On the other hand, {\it coordinatization} is the only way to
individuate the points {\it mathematically} since, as stressed by
Hermann Weyl: ''There is no distinguishing objective property by
which one could tell apart one point from all others in a
homogeneous space: at this level, fixation of a point is possible
only by a {\it demonstrative act} as indicated by terms like {\it
this} and {\it there}.'' (Weyl. 1946, p. 13).

\bigskip

To clarify the situation let us remember that Bergmann and Komar
(1972) gave a passive re-interpretation of active diffeomorphisms
as metric-dependent coordinate transformations $x^{\mu} \mapsto
y^{\mu}(x, {}^4g(x))$ restricted to the solutions of Einstein's
equations (i.e. {\it on-shell}). It can be shown that on-shell
ordinary passive diffeomorphisms and the on-shell Legendre
pull-back of Hamiltonian gauge transformations are two
(overlapping) dense subsets of this set of on-shell
metric-dependent coordinate transformations. Since the Cauchy
surface for the Hole Argument lies outside the hole (where the
active diffeomorphism is the identity), it follows that the
passive re-interpretation of the active diffeomorphism $D^*_A$
must be an on-shell Hamiltonian gauge transformation, so that
Leibniz equivalence is identified with gauge equivalence in the
sense of Dirac constraint theory (${}^4g$ and $D^*_A\, {}^4g$
belong to the same gauge orbit).

\bigskip

What remains to be done is to implement Stachel's suggestion
according to which  the {\it intrinsic pseudo-coordinates} of
Bergmann and Komar (1960) [see also Bergmann (1962) and Komar
(1958)] should be used as individuating fields. These
pseudo-coordinates for $M^4$ (at least when there are no Killing
vectors) are four scalar functions $F^A[w_{\lambda}]$, $A, \lambda
= 1,..,4$, of the four eigenvalues $w_{\lambda}({}^4g,
\partial\, {}^4g)$ of the Weyl tensor. Since these eigenvalues can
be shown to be in general functions of the 3-metric, of its
conjugate canonical momentum (namely of the extrinsic curvature of
$\Sigma_{\tau}$) and of the lapse and shift functions, the
pseudo-coordinates are well defined in phase space and can be used
as a label for the points of $M^4$.

\bigskip

The final step [see Lusanna and Pauri (2005, 2004a,b)] is to
implement the individuation of point-events by considering an
arbitrary admissible 3+1 splitting of $M^4$ with a given time-like
observer and the associated radar 4-coordinates $\sigma^A$ and by
imposing the following gauge fixings to the secondary
super-hamiltonian and super-momentum constraints (the only
restriction on the functions $F^A$ is the orbit condition)

\beq
 \chi^A(\tau ,\sigma^r) = \sigma^A - F^A[w_{\lambda}] \approx 0.
 \nonumber \\
 \eeq

In this way we break completely general covariance and we
determine the gauge variables $\xi^r$ and $\pi_{\phi}$. Then the
$\tau$-constancy of these gauge fixings will produce the gauge
fixings determining the lapse and shift functions. After having
fixed the Lorentz gauge freedom of the cotetrads, we arrive at a
completely fixed gauge in which, after the transition to Dirac
brackets, we get $\sigma^A \equiv {\tilde F}^A[r_{\bar a}(\sigma
), \pi_{\bar a}(\sigma )]$, namely that the radar 4-coordinates of
a point in $M^4_{3+1}$, the copy of $M^4$ coordinatized with the
chosen non-inertial frame, are determined {\it off-shell} by the
four DO of that gauge: in other words the individuating fields are
the genuine tidal effects of the gravitational field. By varying
the functions $F^A$ we can make an analogous off-shell
identification in every other admissible non-inertial frame. The
procedure is consistent, because the DO know the whole 3+1
splitting $M^4_{3+1}$ of $M^4$, being functionals not only of the
3-metric on $\Sigma_{\tau}$, but also of its extrinsic curvature.

\bigskip

Some consequences of this identification of the point-events of
$M^4$ are:\medskip

1) The space-time $M^4$ and the gravitational field are
essentially the same entity. The presence of matter modifies the
solutions of Einstein equations, i.e. $M^4$, but does not play any
role in this identification. Instead matter is fundamental for
establishing a (still lacking) dynamical theory of measurement not
using test objects. As a consequence, instead of the dichotomy
substantivalism/relationism, we believe that this analysis - as a
case study limited to the class of space-times dealt with - may
offer a new more articulated point of view, which can be named
{\it point structuralism} [see also Dorato and Pauri (2004)]. Let
us recall that, in remarkable diversity with respect to the
traditional historical presentation of Newton's absolutism {\it
vis \'a vis} Leibniz's relationism, Newton had a much deeper
understanding of the nature of space and time. In two well-known
passages of {\it De Gravitatione}, Newton expounds what could be
defined an original {\it proto-structuralist view} of space-time.
He writes (our {\it emphasis}):

\begin{quotation}
{\footnotesize \noindent Perhaps now it is maybe expected that I
should define extension as substance or accident or else nothing
at all. But by no means, for it has {\it its own manner of
existence} which fits neither substance nor accidents [\ldots] The
parts of space derive their character from their positions, so
that if any two could change their positions, they would change
their character at the same time and each would be converted
numerically into the other {\it qua} individuals.  The parts of
duration and space are only understood to be the same as they
really are because of their mutual order and positions ({\it
propter solum ordinem et positiones inter se}); nor do they have
any other {\it principle of individuation} besides this order and
position which consequently cannot be altered. (Hall \& Hall,
1962, p.99, p.103.)}
\end{quotation}

\medskip

2) The reduced phase space of this model of general relativity is
the space of abstract DO (pure tidal effects without inertial
effects), which can be thought as four fields on an abstract
space-time ${\tilde M}^4 = \{ equivalence\, class\, of\, all\,
the\, admissible\, non-inertial\, frames\, M^4_{3+1}\,
containing\, the\, associated\, inertial\, effects\}$.

\medskip

3) Each radar 4-coordinate system of an admissible non-inertial
frame $M^4_{3+1}$ has an associated {\it non-commutative
structure}, determined by the Dirac brackets of the functions $
{\tilde F}^A[r_{\bar a}(\sigma ), \pi_{\bar a}(\sigma )]$
determining the gauge.

\medskip

4) Conjecture: there should exist privileged Shanmugadhasan
canonical bases of phase space, in which the DO (the tidal
effects) are also {\it Bergmann observables}, namely
coordinate-independent scalar tidal effects [see Bergmann (1961)].

\bigskip

As a final remark, let us note that these results on the
identification of point-events are {\it model dependent}. In
spatially compact space-times without boundary, the DO are {\it
constants of the motion} due to the frozen picture. As a
consequence, the gauge fixings $\chi^A(\tau ,\sigma^r) \approx 0$
(in particular $\chi^{\tau}$) cannot be used to rebuild the
temporal dimension: probably only the instantaneous 3-space of a
3+1 splitting can be individuated in this way.

\section{Open Problems of Canonical Gravity}

I will finish with a list of the open problems in canonical metric
and tetrad gravity for which there is a concrete hope to be
clarified and solved in the near future.
\bigskip

i) Find a refined Shanmugadhasan canonical transformation allowing
the addition of any kind of matter to the rest-frame instant form
of tetrad gravity. This would allow to study the weak-field
approximation to the two-body problem in a post-Minkowskian
background-independent way by using a Grassmann regularization of
the self-energies, following the track of Crater and Lusanna
(2001) and Alba, Crater and Lusanna (2001). In these papers the
use of Grassmann-valued electric charges to regularize the Coulomb
self-energies allowed to arrive to the Darwin and Salpeter
potentials starting from classical electrodynamics of scalar and
spinning particles, instead of deriving them from quantum field
theory. The solution of the Lichnerowicz equation would allow to
find the expression of the relativistic Newton and
gravito-magnetic action-at-a-distance potentials between the two
bodies (sources, among other effects, of the Newtonian tidal
effects) and the coupling of the particles to the DO of the
gravitational field (the genuine tidal effects) in various radar
coordinate systems: it would amount to a re-summation of the $1/c$
expansions of the Post-Newtonian approximation. Also the
relativistic version of the quadrupole formula for the emission of
gravitational waves from the binary system could be obtained and
some  understanding of how is distributed the gravitational energy
in different coordinate systems could be  obtained. It would also
be possible to study the deviations induced by Einstein's theory
from the Keplerian standards for problems like the radiation
curves of galaxies, whose Keplerian interpretation implies the
existence of dark matter. Finally one could try to define a
relativistic gravitational micro-canonical ensemble generalizing
the Newtonian one developed by Votyakov, Hidmi, De Martino and
Gross (2002).
\bigskip

ii) With more general types of matter (relativistic fluids,
electromagnetic field) it should be possible to develop
Hamiltonian numerical gravity based on the Shanmugadhasan
canonical basis and to study post-Minkowskian approximations based
on power expansions in Newton constant. Moreover one should look
for strong-field approximations to be used in the gravitational
collapse of a ball of fluid.

\bigskip

iii) Find the Hamiltonian formulation of the Newman-Penrose
formalism [see Stewart (1993)], in particular of the 10 Weyl
scalars. Look for the Bergmann observables (the scalar tidal
effects) and try to understand which inertial effects may have a
coordinate-independent form and which are  intrinsically
coordinate-dependent like the ADM energy density. Look for the
existence of a closed Poisson algebra of scalars and for
Shanmugadhasan canonical bases incorporating the Bergmann
observables, to be used to find new expressions for the
super-hamiltonian and super-momentum constraints, hopefully easier
to be solved.

\bigskip

iv) Find all the admissible 3+1 splittings of Minkowski space-time
which avoid the Torre-Varadarajan no-go theorem. Then adapt these
3+1 splittings to tetrad gravity and try to see whether it is
possible to arrive at a multi-temporal background- and coordinate-
independent quantization of the gravitational field, in which only
the Bergmann observables (the scalar tidal effects) are quantized.

\bigskip

v) Try to find the relativistic version of Bell inequalities by
using relativistic particle quantum mechanics in non-inertial
frames.

\vfill\eject

\section{References}

Alba, D., Crater, H. and Lusanna, L. (2001). The Semiclassical
Relativistic Darwin Potential for Spinning Particles in the
Rest-Frame Instant Form: Two-Body Bound States with Spin 1/2
Constituents. {\it International Journal of Modern Physics} {\bf
A16}, 3365-3477 (http://lanl.arxiv.org/abs/hep-th/0103109).
\medskip

Alba,D. and Lusanna, L.(2003). Simultaneity, Radar 4-Coordinates
and the 3+1 Point of View about Accelerated Observers in Special
Relativity (http://lanl.arxiv.org/abs/gr-qc/0311058). \medskip

Alba,D. and Lusanna, L. (2005a). Generalized Radar 4-Coordinates
and Equal-Time Cauchy Surfaces for Arbitrary Accelerated
Observers,  submitted to General Relativity and Gravitation
(http://lanl.arxiv.org/abs/gr-qc/0501090). \medskip

Alba, D. and Lusanna, L. (2005b). Quantum Mechanics in
Non-Inertial Frames with a Multi-Temporal Quantization Scheme: I)
Relativistic Particles (http://lanl.arxiv.org/abs/hep-th/0502060)
to appear in {\it International Journal of Modern Physics}; II)
Non-Relativistic Particles
(http://lanl.arxiv.org/abs/hep-th/0504060).\medskip

Alba, D., Lusanna, L. and Pauri, M. (2005). New Directions in
Non-Relativistic and Relativistic Rotational and Multipole
Kinematics for N-Body and Continuous Systems, invited contribution
for the book {\it Atomic and Molecular Clusters: New Research}
(Nova Science) (http://lanl.arxiv.org/abs/hep-th/0505005).\medskip

Alba, D., Lusanna, L. and Pauri, M. (2004) Multipolar Expansions
for Closed and Open Systems of Relativistic Particles,  {\it
Journal of Mathematical Physics} {\bf 46}, 062505, 1-36
(http://lanl.arxiv.org/abs/hep-th/0402181).\medskip

Alba, D., Lusanna, L. and Pauri, M. (2002). Centers of Mass and
Rotational Kinematics for the Relativistic N-Body Problem in the
Rest-Frame Instant Form. {\it Journal of Mathematical Physics}
{\bf 43}, 1677-1727.\medskip

Agresti,J., De Pietri,R., Lusanna,L. and Martucci,L. (2004).
Hamiltonian Linearization of the Rest-Frame Instant Form of Tetrad
Gravity in a Completely Fixed 3-Orthogonal Gauge: a Radiation
Gauge for Background-Independent Gravitational Waves in a
Post-Minkowskian Einstein Space-Time, {\it General Relativity and
Gravitation} {\bf 36}, 1055-1134;
(http://lanl.arxiv.org/abs/gr-qc/0302084).\medskip .

Arnowitt,R., Deser,S., and Misner,C.W. (1962). The Dynamics of
General Relativity, in L. Witten (ed.), \emph{Gravitation: an
Introduction to Current Research}, (pp. 227--265). NewYork: Wiley.
\medskip

Beig, R. and  O'Murchadha, N. (1987). The Poincare' Group as the
Simmetry Group of Canonical General Relativity, {\it Annals of
Physics (N.Y.)} {\bf 174}, 463-498.\medskip

Bergmann, P.G. and Komar, A. (1960) Poisson Brackets between
Locally Defined Observables in General Relativity, {\it Physical
Review Letters} {\bf 4}, 432-433.\medskip

Bergmann, P.G. (1961) Observables in General Relativity, {\it
Review of Modern Physics} {\bf 33}, 510-514\medskip

Bergmann, P.G. (1962). The General Theory of Relativity, in
S.Flugge (ed.), {\it Handbuch derPhysik}, Vol. IV, {\it Principles
of Electrodynamics and Relativity}, (pp. 247-272). Berlin:
Springer.\medskip

Bergmann,P.G. and Komar, A. (1972), The Coordinate Group
Symmetries of General Relativity, {\it International Journal of
Theoretical Physics}, {\bf 5}, 15-28.\medskip

Crater, H. and Lusanna, L. (2001). The Rest-Frame Darwin Potential
from the Lienard-Wiechert Solution in the Radiation Gauge. {\it
Annals of Physics (N.Y.)} {\bf 289}, 87-177
(http://lanl.arxiv.org/abs/hep-th/0001046).\medskip

Christodoulou, D., and Klainerman, S. (1993). {\it The Global
Nonlinear Stability of the Minkowski Space}. Princeton: Princeton
University Press.\medskip

De Pietri,R., Lusanna,L., Martucci,L. and Russo,S. (2002). Dirac's
Observables for the Rest-Frame Instant Form of Tetrad Gravity in a
Completely Fixed 3-Orthogonal Gauge, {\it Geneneral Relativity and
Gravitation} {\bf 34}, 877-1033;
(http://lanl.arxiv.org/abs/gr-qc/0105084).\medskip

DeWitt,B. (1967) Quantum Theory of Gravity, I) The Canonical
Theory, {\it Physical Review} {\bf 160}, 1113-1148. II) The
Manifestly Covariant Theory, {\bf 162}, 1195-1239.\medskip

Dirac, P.A.M. (1949). Forms of Relativistic Dynamics, {\it Review
of Modern Physics} {\bf 21}, 392-399.\medskip

Dirac, P.A.M. (1955). Gauge Invariant Formulation of Quantum
Electrodynamics, {\it Canadian Journal of Physics} {\bf 33},
650-659.\medskip

Dorato.M and Pauri,M. (2004) Holism and Structuralism in Classical
and Quantum General Relativity, Pittsburgh-Archive, ID code 1606,
forthcoming in (2006). S.French and D.Rickles (eds.), {\it
Structural Foundations of Quantum Gravity}, Oxford: Oxford
University Press.\medskip

Earman,J. and Norton,J. (1987). What Price Spacetime
Substantivalism? The Hole Story, \emph{British Journal for the
Philosophy of Science} \textbf{38}, 515--525.\medskip

Einstein,A. (1914). Die formale Grundlage der allgemeinen
Relativit\"atstheorie, in \emph{Preuss. Akad. der Wiss. Sitz.},
(pp. 1030--1085).\medskip

Einstein,A. (1916). Die Grundlage der allgemeinen
Relativit\"atstheorie, \emph{Annalen der Physik} \textbf{49},
769--822; (1952) translation by W. Perrett and G. B. Jeffrey, The
Foundation of the General Theory of Relativity, in \emph{The
Principle of Relativity}, (pp. 117--118). New York: Dover.
\medskip

Friedrich, H. and Rendall, A. (2000). The Cauchy Problem for
Einstein Equations, in B.G.Schmidt (ed.), {\it Einstein's Field
Equations and their Physical Interpretation}. Berlin: Springer;
(http://lanl.arxiv.org/abs/gr-qc/0002074).\medskip

Hall, A.R. and Hall, M.B., (eds.), (1962). {\it De Gravitatione et
Aequipondio Fluidorum, Unpublished Scientific Papers of Isaac
Newton. A Selection from the Portsmouth Collection in the
University Library}. Canbridge: Cambridge University Press.
\medskip

Hawking, W.S. and Horowitz, G.T. (1996), The Gravitational
Hamiltonian, Action, Entropy and Surface Terms, {\it Classical and
Quantum Gravity} {\bf 13}, 1487-1498.\medskip

Hilbert,D. (1917) Die Grundlagen der Physik. (Zweite Mitteilung),
\emph{Nachrichten von der K\"oniglichen Gesellschaft der
Wissenschaften zu G\"ottingen, Mathematisch-physikalische Klasse},
(pp. 53--76).\medskip

Komar,A. (1958). Construction of a Complete Set of Independent
Observables in the General Theory of Relativity, {\it Physical
Review} {\bf 111}, 1182-1187.\medskip

Kuchar, K. (1992). Time and Interpretations of Quantum Gravity, in
{\it Proceedings of the 4th Canadian Conference on General
Relativity and Relativistic Astrophysics}, (pp. 211-314).
Singapore: World Scientific.\medskip

Kuchar, K. (1993). Canonical Quantum Gravity. In G.Kunstatter,
D.E.Vincent, and J.G.Williams (eds.), {\it Cordoba 1992, General
Relativity and Gravitation}, (pp.119-150);
(http://lanl.arxiv.org/abs/gr-qc/9304012).\medskip

Lichnerowicz, A. (1964). Propagateurs, Commutateurs et
Anticommutateurs en Relativite Generale, in {\it Relativity,
Groups and Topology}, Les Houches 1963, eds. DeWitt, C. and
DeWitt, B., New York, Gordon and Breach.\medskip

Lusanna, L. (1993). The Shanmugadhasan Canonical Transformation,
Function Groups and the Second Noether Theorem, {\it International
Journal of Modern Physics} {\bf A8}, 4193-4233.\medskip

Lusanna, L. (1995). Classical Yang-Mills Theory with Fermions. I.
General Properties of a System with Constraints, {\it
International Journal of Modern Physics} {\bf A10}, 3531-3579; II.
Dirac's Observables, {\it International Journal of Modern Physics}
{\bf A10}, 3675-3757.\medskip

Lusanna, L. (1997). The N- and 1-Time Classical Description of
N-Body Relativistic Kinematics and the Electromagnetic
Interaction, {\it International Journal of Modern Physics} {\bf
A12}, 645-722.\medskip

Lusanna, L. (2001). The Rest-Frame Instant Form of Metric Gravity,
{\it General Relativity and Gravitation} {\bf 33}, 1579-1696;
(http://lanl.arxiv.org/abs/gr-qc/0101048).\medskip

Lusanna, L. (2004). The Chronogeometrical Structure of Special and
General Relativity: towards a Background-Independent Description
of the Gravitational Field and Elementary Particles, invited
contribution to the book {\it Progress in General Relativity and
Quantum Cosmology} (Nova Science)
(http://lanl.arxiv.org/abs/gr-qc/0404122).\medskip

Lusanna,L. and Russo,S. (2002) A New Parametrization for Tetrad
Gravity, {\it General Relativity and Gravitation} {\bf 34},
189-242; (http://lanl.arxiv.org/abs/gr-qc/0102074).\medskip

Lusanna, L. and Pauri, M. (2005) General Covariance and the
Objectivity of Space-Time Point-Events, talk at the Oxford
Conference on Spacetime Theory (2004), to appear in {\it History
and Philosophy of Modern Physics}
(http://lanl.arxiv.org/abs/gr-qc/0503069).\medskip

Lusanna, L. and M.Pauri, M. (2004a) The Physical Role of
Gravitational and Gauge Degrees of Freedom in General Relativity.
I: Dynamical Synchronization and Generalized Inertial Effects, to
appear in {\it General Relativity and Gravitation};
(http://lanl.arxiv.org/abs/gr-qc/0403081).\medskip

Lusanna,L. and Pauri,M. (2004b). The Physical Role of
Gravitational and Gauge Degrees of Freedom in General Relativity.
II: Dirac versus Bergmann Observables and the Objectivity of
Space-Time, to appear in {\it General Relativity and Gravitation};
(http://lanl.arxiv.org/abs/gr-qc/0407007).\medskip

Mashhoon, B. (1990). The Hypothesis of Locality in Relativistic
Physics, {\it Physics Letters} {\bf A145}, 147-153.\medskip

Mashhoon, B. and Muench, U.(2002). Length Measurement in
Accelerated Systems, {\it Annalen der Physik (Leipzig)} {\bf 11},
532-547. \medskip

Mashoon, B. (2003). The Hypothesis of Locality and its Limitations
(http://lanl.arxiv.org/abs/gr-qc/0303029).\medskip

M$\o$ller, C. (1957). The Theory of Relativity. Oxford, Oxford
Univeersity Press.\medskip

Moncrief, V. (1979). Gribov Degeneracies: Coulomb Gauge Conditions
and Initial Value Constraints, {\it Journal of Mathematical
Physics} {\bf 20}, 579-585.\medskip

Nicolai, H., Peeters, K. and Zamaklar, M.(2005). Loop Quantum
Gravity: an Outside View
(http://lanl.arxiv.org/abs/hep-th/0501114).\medskip

Norton, J. (2005). Einstein's Investigations of Galilean Covariant
Electrodynamics prior to 1905,
(http://philsci-archive.pitt.edu/archive/00001743/)\medskip

Norton,J.(1993). General Covariance and the Foundations of General
Relativity: Eight Decades of Dispute, {\it Rep.Prog.Phys.} {\bf
56}, 791-858.\medskip

Regge, T. and Teitelboim, C. (1974). Role of Surface Integrals in
the Hamiltonian Formulation of General Relativity, {\it Annals of
Physics (N.Y.)} {\bf 88}, 286-318.\medskip

Rendall, A. (1998). Local and Global Existence Theorems for the
Einstein Equations, {\it Online Journal Living Reviews in
Relativity} {\bf 1}, n. 4; {\it ibid}. (2000) {\bf 3}, n. 1;
(http://lanl.arxiv.org/abs/gr-qc/0001008).\medskip

Shanmugadhasan,S. (1973). Canonical Formalism for Degenerate
Lagrangians, {\it Journal of Mathematical Physics} {\bf 14},
677-687.\medskip

Smolin, L. (2003). How Far are we from Quantum Gravity?
(http://lanl.arxiv.org/abs/hep-th/0303185).\medskip

Stachel,J. (1980). Einstein's Search for General Covariance,
1912--1915. {\it Ninth International Conference on General
Relativity and Gravitation}, Jena, ed. E.Schmutzer (Cambridge
Univ.Press, Cambridge, 1983).\medskip

Stewart,J. (1993). {\it Advanced General Relativity}, Cambridge:
Cambridge University Press.\medskip

Teitelboim, C. (1980). The Hamiltonian Structure of Spacetime, in
{\it General Relativity and Gravitation}, ed. Held, A., Vol. I,
Plenum, New York.\medskip

Torre, C.G. and Varadarajan, M. (1999) Functional Evolution of
Free Quantum Fields, {\it Classical and Quantum Gravity} {\bf 16},
2651-2668.\medskip

Votyakov, E.V., Hidmi, H.I., De Martino, A. and Gross, D.H.E.
(2002).Microcanonical Mean-Field Thermodynamics of
Self-Gravitating and Rotating Systems, {\it Physical Review
Letters} {\bf 89}, 031101
(http://lanl.arxiv.org/abs/cond-mat/0202140).\medskip

Wald,R.M. (1984) {\it General Relativity}. Chicago: University of
Chicago Press.\medskip

Weyl,H.,(1946). Groups, Klein's Erlangen Program. Quantities,
ch.I, sec.4 of {\it The Classical Groups, their Invariants and
Representations}, 2nd ed., (pp.13-23). Princeton: Princeton
University Press.

\end{document}